\documentclass[showpacs,showkeys,prl,twocolumn,final]{revtex4}%
\usepackage{amsfonts}
\usepackage{amsmath}
\usepackage{amssymb}
\usepackage{graphicx}%
\providecommand{\U}[1]{\protect\rule{.1in}{.1in}}

\begin{document}
\preprint{ }
\title[Andreev in BECs]{Andreev reflection in bosonic condensates}
\author{I. Zapata}
\author{F. Sols}
\affiliation{Departamento de F\'{\i}sica de Materiales, Universidad Complutense de Madrid,
E-28040 Madrid, Spain}

\begin{abstract}
We study the bosonic analog of Andreev reflection at a normal-superfluid
interface where the superfluid is a boson condensate. We model the normal
region as a zone where nonlinear effects can be neglected. Against the
background of a decaying condensate, we identify a novel contribution to the
current of reflected atoms. The group velocity of this Andreev reflected
component differs from that of the normally reflected one. For a
three-dimensional planar or two-dimensional linear interface Andreev reflection is neither specular nor conjugate.

\end{abstract}
\volumeyear{2008}
\volumenumber{number}
\issuenumber{number}
\eid{identifier}
\startpage{1}
\endpage{ }
\maketitle

Andreev reflection occurs at a normal-superconductor interface when an
incident electron/hole on the normal side is reflected as a hole/electron
\cite{Andreev64}. It is a current-carrying process which, within a mean-field
description, conserves the quasiparticle but not the charge current. The
Andreev reflection of an incident hole is equivalent to the emission of a
singlet electron pair from the superconductor into the normal metal
\cite{2e-picture}. It has been observed in superconductors \cite{scr-exp} and
superfluid $^{3}$He \cite{Enrico93}, and has been proposed for fermionic
quantum gases \cite{Schaeybroeck07}. In these Fermi superfluids the conversion
between particles and holes reflects the fermion-pair structure of the
superfluid phase. Andreev reflection determines the transport properties of fermionic superfluid-normal interfaces. It seems natural to investigate its bosonic analog. Andreev-like processes have been proposed for cold atom
realizations of Bose-Hubard models \cite{Daley08} and Luttinger liquids
\cite{Tokuno07}. These two works investigate the generation of reflected dips
in the particle density following the incidence of a density bump on an interface.

Here we investigate the possible analog of Andreev reflection at the interface
between a bosonic superfluid and a normal gas. We seek an analog closer to the
fermionic case by exploring the conversion between different types of
Bogoliubov quasiparticles, which for bosons are defined through the
transformation $\psi=e^{i\phi}\sum_{n}(u_{n}\gamma_{n}-v_{n}^{\ast}\gamma
_{n}^{+})$, where $\psi$ is the field fluctuation operator, $\phi$ is the condensate phase, and $\gamma_{n}$
destroys quasiparticle $n$ \cite{Fetter72}. In bosonic fields the $u$ and $v$
components play asymmetric roles;\ for instance, under usual circumstances
$v=0$ in a normal gas. In this context, we define Andreev reflection in a
normal-superfluid (NS) interface as the scattering event which generates a
quasiparticle with $v\neq0$ on the N side following the incidence of a
conventional ($v=0$) quasiparticle from the same side. The hole-like
quasiparticle (which however contributes positively to the density) travels at
a speed different from the normally reflected one, which should be
experimentally observable.

A natural way of forming a NS interface in boson gases is to create a
potential step that forces a difference in the atom densities on each side of
the step. At a given temperature, the densities may be such that the gas is
normal on the left and superfluid on the right. However, since in a normal
boson gas the chemical potential stays below zero kinetic energy, transport
near equilibrium is possible only if the condensate is confined. The $v$
component represents propagation below the chemical potential, so it can only
be evanescent on the N side within that scenario \cite{Wynveen00,Poulsen03}.
As Andreev reflection from a confined condensate is ruled out, one is led to
consider the alternative case of a decaying condensate.

We define the normal (or non-superfluid) region as that in which the
condensate flows faster than the local speed of sound;\ in our case, the
outgoing coherent beam. For simplicity, we assume the interaction coupling
strength to be zero on the N side, which does not change the essential
physics. In general, transport through an interface separating subsonic from
supersonic flow poses a new paradigm in superfluid transport \cite{Leboeuf01}.

Like for fermions, particle current is not conserved within a
(non-selfconsistent) mean-field description of Andreev processes
\cite{SolsA94,Sanchez97}. In both cases the condensate is responsible for the
loss or gain of particles. However, there are several important differences
with respect to the fermion case. The eigenvalue problem whose solutions are
the Bogoliubov quasiparticles is non-Hermitian. This results in a peculiar
normalization condition [$\nu\equiv\int(|u|^{2}-|v|^{2})dx=1$] for the
quasiparticle wave functions \cite{Pethick02}. We will see that the
conservation of $\nu$ yields unconventional relations between the scattering
amplitudes. A counterintuitive but straightforward consequence is that, unlike
for fermions, both particle-like and hole-like excitations carry an increase
in the density (with respect to the Bogoliubov vacuum), since $\rho\sim
|u|^{2}+|v|^{2}$. The existence of these differences suggests that bosonic
quasiparticles define a rich novel class of quantum transport problems.

The need for a confined condensate to generate Andreev reflection can be
better appreciated from a study of the Bogoliubov -- de Gennes (BdG) equations
for bosons \cite{Dalfovo99,Leggett01}:%

\begin{figure}
[ptb]
\begin{center}
\includegraphics[
height=2.5944in,
width=3.5907in
]%
{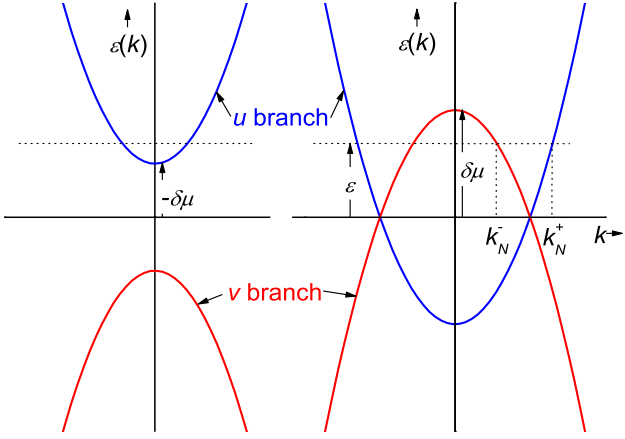}%
\caption{
Dispersion relation on the normal side for branches $u$ (particle-like) and $v$ (hole-like) for a confined (left) and a decaying (right) condensate. Labels are explained after Eq. (\ref{eqnBdG}) in main text. Elastic conversion between propagating branches (Andreev reflection) requires $\delta \mu >0$, i.e. a decaying condensate.
}%
\label{grPlotNDispersionRelation}%
\end{center}
\end{figure}

%

\begin{equation}%
\begin{array}
[c]{c}%
i\partial_{t}\dbinom{u}{v}=\left(
\begin{array}
[c]{cc}%
\mathcal{L}(x) & -n_{0}(x)g(x)\\
n_{0}(x)g(x) & -\mathcal{L}^{\ast}(x)
\end{array}
\right)  \dbinom{u}{v}\\
\\
\mathcal{L}(x)\equiv-\frac{1}{2}D_{x}^{2}+V(x)+2n_{0}(x)g(x)-\mu
\end{array}
\label{eqnBdG}%
\end{equation}
Here, $\hbar=m=1$ units have been used, $n_{0}(x)$ is the condensate density,
$g(x)$ is the (piecewise constant) coupling strength, $D_{x}\equiv\partial
_{x}-i\phi^{\prime}$ with $\phi(x)$ the condensate phase, and $\mu$ is the
condensate chemical potential. The scattering channels are defined by the
plane wave solutions ($u,v\sim e^{i(kx-\varepsilon t)}$) in the flat region
$V(x)=V_{0}$. The dispersion relations in the non-superfluid region
($g(x)n_{0}(x)=0$) are shown in Fig. \ref{grPlotNDispersionRelation}. In the
confined case ($\delta\mu\equiv\mu-V_{0}<0$), the left Fig.
\ref{grPlotNDispersionRelation} shows that, away from the interface,
asymptotic propagation with $\varepsilon>0$ is only possible for the $u$
(particle-like) component, while the $v$ (hole-like) component can only be
evanescent. A decaying ($\delta\mu>0$) condensate exhibits a richer scenario.
The right Fig. \ref{grPlotNDispersionRelation} shows a crossing of the $u$-
and $v$-branches. This permits elastic conversion from a particle into a
\textquotedblleft hole\textquotedblright, in formal analogy with fermionic
Andreev reflection. For a given energy within the allowed crossing range
$\varepsilon\in(-\delta\mu,\delta\mu)$, there are four propagating waves. The
right (left) going wave numbers are $k_{N}^{+}$ and $-k_{N}^{-}$
\ ($-k_{N}^{+}$ and $k_{N}^{-}$), where $k_{N}^{\pm}\equiv\sqrt{2(\delta\mu
\pm\varepsilon)}$. The group velocities are those of a quadratic dispersion
relation, $w_{N}^{\pm}=k_{N}^{\pm}$. Thus, for an incident particle from the
normal side with velocity $w_{i}=\sqrt{2(\delta\mu+\varepsilon)}$, there are
two outgoing channels with velocity $w_{f}$ satisfying $w_{f}=w_{i}$ (normal
reflection) or $w_{i}^{2}+w_{f}^{2}=4~\delta\mu$ (Andreev reflection). The
prediction of this novel form of propagation on the normal side is the main
result of this paper.

On the superfluid side, excitations travel with wavevector $k_{S}^{+}%
\equiv\sqrt{2}[\sqrt{\varepsilon^{2}+\mu^{2}}-\mu]^{1/2}$. Here, $\mu=gn_{+}$,
with $n_{\pm}\equiv\lim_{x\rightarrow\pm\infty}n_{0}(x)$. At low energies, the
speed $w_{S}=d\varepsilon/dk_{S}^{+}$ approaches the speed of sound $c=gn_{+}%
$. An Andreev transmission process occurs for energies $\varepsilon
\in(0,\delta\mu)$ when a phonon incident from S is transmitted into N as a
hole-like ($u=0$) quasiparticle. The other transmission channel is
particle-like ($v=0$) and travels in N at a different velocity ($w_{N}^{+}$
vs. $w_{N}^{-}$). Remarkably, the outgoing hole-like scattering channels which
characterize an Andreev process may exist despite its anomalous normalization
$\nu<0$. We note, however, that the conservation of $\nu$ applies to the total
scattering state: a particular channel may have $\nu<0$ whereas the total
state retains the standard normalization $\nu>0$.

\bigskip

\bigskip%

\begin{figure}
[ptb]
\begin{center}
\includegraphics[
height=2.9974in,
width=3.6452in
]%
{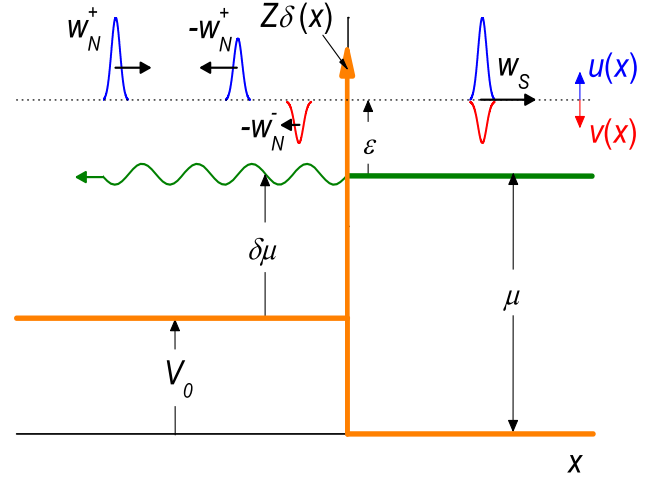}%
\caption{Schematic plot of the scattering problem. Against the background of
the coherent beam emitted by a decaying condensate, a single atom impinges
from the normal side. Three scattering processes result:\ normal reflection,
transmission into a phonon mode on the superfluid side, and Andreev reflection
on the normal side, with group velocities $-w_{N}^{+},w_{S},-w_{N}^{-}$,
respectively. Labels are explained after Eq. (\ref{eqnBdG}) in main text.}%
\label{grPotentialPlot}%
\end{center}
\end{figure}

Bogoliubov quasiparticles are generally defined against the background of a
particular condensate wave function. We consider a one-atom potential of the
type shown in Fig. \ref{grPotentialPlot}, which provides a model for a leaking
condensate. The condensate wave function $\Psi$ satisfies the time-dependent
Gross-Pitaevskii equation%

\begin{align}
i\partial_{t}\Psi &  =-\tfrac{1}{2}\partial_{xx}\Psi+V(x)\Psi+g(x)|\Psi
|^{2}\Psi~,\label{eqnGrossPitaevskii}\\
V(x)  &  =Z\,\delta(x)+V_{0}\,\theta(-x)~. \label{eqnPotentialProfile}%
\end{align}

In the asymptotic regions it admits a solution of the type $\Psi
(x\rightarrow\pm\infty)\sim e^{i(qx-\mu t)}$. In the previous discussion on
scattering channels, we have assumed $q=0$ for simplicity, which ---it can be
shown--- does not change the main conclusions.

A stationary solution to (\ref{eqnGrossPitaevskii}) and
(\ref{eqnPotentialProfile}) can be found analytically \cite{details}. An
important question is whether or not that solution is stable. For $Z/\sqrt
{\mu}\gg1$, the exact stationary solution resembles closely the wave function
$\Psi(x)=\sqrt{n_{-}}\theta(-x)+\sqrt{n_{+}}\theta(x)$. For this approximate,
step-like condensate, we have proved that the solutions to the BdG equations
(\ref{eqnBdG}) do not have complex eigenvalues. In other words, in the thick
barrier limit the exact stationary solution resembles a solution which has
been proven to be stable. In general, the condensed atom density on the N side
is uniform and equal to $n_{-}=|j|/\sqrt{2\delta\mu}$, with atom velocity
$w=-\sqrt{2\delta\mu}$, as determined by simple energy considerations. In this
thick barrier limit, the resulting current is $j\simeq-\mu^{2}\sqrt{2\delta
\mu}/gZ^{2}$. This solution applies to an interface separating a supersonic
(normal) from a subsonic (superfluid) region \cite{Leboeuf01}.

Next we write the general form of the scattering states within the
approximation of a flowless condensate. For an atom incident from the N side
with $|\varepsilon|<\delta\mu$ we have%

\begin{equation}
\psi_{N}\equiv\left\{
\begin{array}
[c]{ll}%
\binom{1}{0}\left[  \chi_{N}^{+}+r_{n}\left(  \chi_{N}^{+}\right)  ^{\ast
}\right]  +\binom{0}{1}r_{a}\chi_{N}^{-}, & x\rightarrow-\infty\\
\binom{u_{0}}{v_{0}}t_{p}\chi_{S}+\binom{-v_{0}}{u_{0}}t_{e}~e^{-k_{S}^{-}%
~x}, & x\rightarrow\infty
\end{array}
\right.  \label{eqnScatteringFromTheLeft}%
\end{equation}
where $\chi_{N}^{\pm}\equiv e^{ik_{N}^{\pm}x}/\sqrt{w_{N}^{\pm}}$, $\chi
_{S}\equiv e^{i\mathrm{sgn}(\varepsilon)k_{S}^{+}x}/\sqrt{w_{S}}$ and
$k_{S}^{-}\equiv\sqrt{2}[\sqrt{\varepsilon^{2}+\mu^{2}}+\mu]^{1/2}$
and~$u_{0},v_{0}$ are such that $u_{0}^{2}+v_{0}^{2}=\sqrt{\varepsilon^{2}%
+\mu^{2}}/|\varepsilon|$ and $u_{0}/v_{0}=\sqrt{1+\varepsilon^{2}/\mu^{2}%
}+\varepsilon/\mu$. For an incoming phonon from the S side with energy
$0<\varepsilon<\delta\mu$, we have%

\begin{equation}
\psi_{S}\equiv\left\{
\begin{array}
[c]{ll}%
\binom{1}{0}t_{n}\left(  \chi_{N}^{+}\right)  ^{\ast}+\binom{0}{1}t_{a}%
\chi_{N}^{-}, & x\rightarrow-\infty,\\
\binom{u_{0}}{v_{0}}\left[  \chi_{S}^{\ast}+r_{p}\chi_{S}\right]
+\binom{-v_{0}}{u_{0}}r_{e}~e^{-k_{S}^{-}~x}, & x\rightarrow\infty,
\end{array}
\right.  \label{eqnScatteringFromTheRight}%
\end{equation}
The subindeces stand for normal ($a$), Andreev ($n$), phonon ($p$) and
evanescent ($e$), while $r$ and $t$ represent reflection and transmission
coefficients. Equations (\ref{eqnScatteringFromTheLeft}) and
(\ref{eqnScatteringFromTheRight}) do not include the case $\varepsilon
>\delta\mu$, which shows no Andreev processes, or that of an incoming channel
with $\nu<0$, which does not seem realizable unless the beam comes from
another Bose condensate.

Conservation laws may be obtained from the generalized Wronskian
\begin{equation}
W_{ij}\equiv u_{i}D_{x}^{\ast}u_{j}^{\ast}-u_{j}^{\ast}D_{x}u_{i}+v_{i}%
D_{x}v_{j}^{\ast}-v_{j}^{\ast}D_{x}^{\ast}v_{i}~, \label{eqnWronskian}%
\end{equation}
which is constant for any two stationary solutions of Eq. (\ref{eqnBdG}) with
same energy $\varepsilon$. When applied to $\psi_{N}$ and $\psi_{S}$, the
constancy of (\ref{eqnWronskian}) generates the following relations between
the scattering amplitudes:%

\begin{equation}%
\begin{array}
[c]{c}%
|r_{n}|^{2}-|r_{a}|^{2}+\mathrm{sgn}(\varepsilon)~|t_{p}|^{2}=|t_{n}%
|^{2}-|t_{a}|^{2}+|r_{p}|^{2}=1\\
r_{n}~t_{n}^{\ast}+r_{p}^{\ast}~t_{p}-r_{a}~t_{a}^{\ast}=0
\end{array}
\label{eqnWronskianEqualities}%
\end{equation}
The negative signs here can be traced back to the non-Hermitian character of
the effective Hamiltonian in the bosonic BdG equations (\ref{eqnBdG}). This
represents a major difference with respect to the fermionic Bogoliubov problem
and is a source of difficulties in the interpretation of the particle-hole
transformation for bosons.

If we assume a step-like profile for the condensate density, the scattering
amplitudes can be found analytically. Here we note some important trends in
the low transparency limit ($Z/\sqrt{\mu}\gg1$):\ The Andreev reflection
amplitude ${|{{r_{a}}}|}$ scales like $Z^{-2}$, which indicates that, like for
superconductors \cite{Blonder82}, an Andreev reflection process involves the
transmission of two atoms. This reflects the pairing correlations which exist
in the quantum depletion cloud of the condensate and which are expressed
through the structure of the Bogoliubov transformation and the BdG equations
\cite{Leggett01}. Analogously, the scaling $|t_{a}|,|t_{p}|\sim Z^{-1}$
suggests that for large $Z$ the corresponding scattering processes involve the
transmission of one atom.

More information on conservation laws can be obtained from the continuity equation%

\begin{equation}
\partial_{t}|u|^{2}+\partial_{x}j_{u}=\partial_{t}|v|^{2}+\partial
_{x}j_{v^{\ast}}=2~n_{0}(x)g(x)\operatorname{Im}(uv^{\ast})~,
\label{eqnContinuity}%
\end{equation}
where $j_{s}\equiv\operatorname{Im}(s^{\ast}D_{x}s)$, which tells us
simultaneously about the lack of conservation of the atom mass current
(density $|u|^{2}+|v|^{2}$, current $j_{u}+j_{v^{\ast}}$) and the conservation
of the quasiparticle current (density $|u|^{2}-|v|^{2}$, current
$j_{u}-j_{v^{\ast}}$), the latter being also expressed by the constancy of the
Wronskian (\ref{eqnWronskian}). Both mass density and current are defined as
their total values in the presence of the excitation of wave function
$\binom{u}{v}$ minus the corresponding values of the Bogoliubov vacuum, which
includes the zero-temperature depletion cloud \cite{Fetter72} and thus a
fluctuating atom flux. We note that, while the $u$ component carries the
current in the usual way, the $v$ does it in the opposite one. Another
important feature is that, unlike for its fermionic counterpart
\cite{Blonder82}, the mass current has the same sign as the group velocity in
all channels (the mass density is always positive).

As in the case of superconductivity \cite{Sanchez97,Blonder82}, we interpret
the non-conservation of atom number as the exchange of atoms with the
Bogoliubov vacuum. The net balance of extracted atoms per unit time can be
calculated by substracting, from the current at a point deep in the superfluid
region, the current at any point in the normal region, $\Delta I\equiv
j(x_{+})-j(x_{-})$. We obtain%

\begin{equation}
\Delta I=|r_{n}|^{2}+|r_{a}|^{2}+\frac{|t_{p}|^{2}}{u_{0}^{2}+v_{0}^{2}%
}-1\simeq\frac{-2~\mathrm{sgn}(\varepsilon)~k_{N}^{+}{}~k_{S}^{+}%
}{1+\varepsilon^{2}/\mu^{2}}\frac{1}{Z^{2}}~. \label{eqnCurrentBalance}%
\end{equation}
For the second equality, which gives the current to leading order in $Z^{-1}$,
we have used (\ref{eqnWronskianEqualities}) and the exact results for the
scattering amplitudes \cite{details}. Thus, for an atom incident from the N
side, the scattering process which follows is such that atoms are lost to the
Bogoliubov vacuum if $\varepsilon>0$, while they are extracted from that
vacuum if $\varepsilon<0$. From the continuity equation (\ref{eqnContinuity})
it is clear that the transfer of atoms between the condensate and the
quasiparticle field takes place exclusively on the S side, since $g(x)=0$ in N.%

\begin{figure}
[ptb]
\begin{center}
\includegraphics[
height=2.9776in,
width=2.7415in
]%
{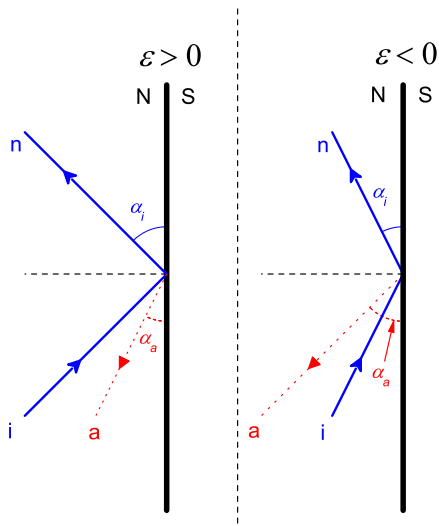}%
\caption{Andreev reflection at a bosonic three-dimensional NS\ interface.
Indeces $i,n,a$ stand for incoming, Andreev reflected and normally reflected
channels. The arrows represent group velocities. The angle of the Andreev
reflection channel is sensitive to the sign of the energy $\varepsilon$.}%
\label{grAndreev3D}%
\end{center}
\end{figure}

The generalization of the above discussion to the case of a three-dimensional planar or two-dimensional linear interface yields qualitative differences with respect to the superconducting case. Invoking the conservation of energy and parallel
momentum, we find that for $\varepsilon\in(-\delta\mu,\delta\mu)$ the beam
angles (see Fig. \ref{grAndreev3D}) satisfy the relation%
\begin{equation}
(1-\varepsilon/\delta\mu)\tan^{2}\alpha_{i}-(1+\varepsilon/\delta\mu)\tan
^{2}\alpha_{a}=2~\varepsilon/\delta\mu~. \label{angles}%
\end{equation}
This result leads to two conclusions: (a) If $0<\varepsilon<\delta\mu$, \ no
Andreev reflection occurs for $\tan\alpha_{i}<\sqrt{\varepsilon/2(\delta
\mu-\varepsilon)}$ and one has $\alpha_{a}<\alpha_{i}$. (b) If $-\delta
\mu<\varepsilon<0$ Andreev reflection is possible, but only satisfying
$\alpha_{a}>\alpha_{i}$ and $\tan\alpha_{a}\geq\sqrt{-\varepsilon/2(\delta
\mu+\varepsilon)}$.

The observation of bosonic Andreev reflection poses a challenge. A tunable
coupling constant can be achieved through the handling of Feshbach resonances
\cite{Fattori08} or, in the 1D limit, by increasing the width of the confining
channel. Such long channels can form in the vicinity of a planar semiconductor
chip. A potential barrier at the interface can be introduced with a properly
focused, blue-detuned laser. The crucial measurement of escaping velocities may rely on precise Bragg spectroscopy or on the interference between incoming and outgoing beams. The 2D/3D case presents greater potential flexibility because anomalous reflection can occur in a range of different directions. Confinement by a linear interface in 2D could be achieved with a laser beam.
For instance, $^{87}$Rb atoms may be confined in a waveguide with
transverse trapping frequency of 500 Hz and experience a delta-barrier of
$Z=5\sqrt{\mu}$ if they are exposed to a narrow Gaussian laser of 1 $\mu$m
waist, blue-detuned by 10 nm to the 5$^{2}$S$_{1/2}$-5$^{2}$P$_{1/2}$
transition \cite{Metcalf99} (780.24 nm, linewidth 12$\pi$ Hz, saturation
intensity 1.64 mW cm$^{-2}$). With a step barrier such that $\delta\mu=0.5\mu
$, bulk superfluid linear densities in the range 0.2--20 $\mu$m$^{-1}$ place
the system in the low-density mean-field region, where the present estimate
applies. For that density range we obtain 1D coherence-length 7--0.7 $\mu$m,
laser intensity 80--800 mW cm$^{-2}$, and 10--1000 $\mu$m s$^{-1}$ for the
Andreev velocity scale ($\sim2\sqrt{\delta\mu}$), close to the superfluid
sound speed of 70--700 $\mu$m s$^{-1}$. For these parameters the emitted
condensate linear density is 0.04 times the bulk S density, while the
corresponding condensate escape velocity is the same as the speed of sound in
S, the equality being due to our choice of $\delta\mu$.

We acknowledge valuable discussions with J. Brand, N. Davidson, E. Demler,
S. Foelling, A. J. Leggett, C. Lobo, H. Michinel, E. Muller, and G. Shlyapnikov.
This work has
been supported by the Ram\'{o}n Areces Foundation and by MEC (Spain) through
Grant FIS2007-65723. The authors thank the hospitality and support of the
Institut Henri Poincare - Centre Emile Borel, where part of this work was done.


\bigskip

\begin{thebibliography}{99}                                                                                               %


\bibitem {Andreev64}A.~F. Andreev, Sov. Phys. JETP \textbf{19}, 1228 (1964).

\bibitem {2e-picture}P.~Samuelsson \textit{et al.}, Phys. Rev. Lett.
\textbf{91}, 157002 (2003); New J. Phys. \textbf{7}, 176 (2005);\ E.~Prada and
F.~Sols, Eur. Phys. J. B \textbf{40}, 379 (2004).

\bibitem {scr-exp}S. I. Bozhko \textit{et al.}, JETP Lett. \textbf{36}, 153
(1982); P. A. M. Benistant \textit{et al.}, Phys. Rev. Lett. \textbf{51}, 817
(1983); G. E. Blonder and M. Tinkham, Phys. Rev. B \textbf{27}, 112 (1983).

\bibitem {Enrico93}M.~P. Enrico \textit{et al.}, Phys. Rev. Lett. \textbf{70},
1846 (1993).

\bibitem {Schaeybroeck07}B.~Van Schaeybroeck and A. Lazarides, Phys. Rev.
Lett. \textbf{98}, 170402 (2007).

\bibitem {Daley08}A.~J. Daley \textit{et al.}, Phys. Rev. Lett. \textbf{100},
110404 (2008).

\bibitem {Tokuno07}A.~Tokuno \textit{et al.}, Phys. Rev. Lett. \textbf{100},
140402 (2008).

\bibitem {Fetter72}A.~L. Fetter, Ann. Phys. (New York) \textbf{70}, 67 (1972).

\bibitem {Wynveen00}A.~Wynveen \textit{et al.}, Phys. Rev. A \textbf{62},
023602 (2000).

\bibitem {Poulsen03}U. V. Poulsen and K. M\o {}lmer, Phys. Rev. A \textbf{67},
013610 (2003).

\bibitem {Leboeuf01}P. Leboeuf and N. Pavloff, Phys. Rev. A \textbf{64},
033602 (2001).

\bibitem {SolsA94}F.~Sols and J.~Ferrer, Phys. Rev. B \textbf{49}, 15913 (1994).

\bibitem {Sanchez97}J.~S\'{a}nchez-Ca\~{n}izares and F.~Sols, Phys. Rev. B
\textbf{55}, 531 (1997).

\bibitem {Pethick02}C. J. Pethick and H.~Smith, \emph{Bose-Einstein
Condensation in Dilute Gases} (Cambridge University Press, Cambridge, 2002).

\bibitem {Dalfovo99}F.~Dalfovo \textit{et al.}, Rev. Mod. Phys. \textbf{71},
463 (1999).

\bibitem {Leggett01}A.~J. Leggett, Rev. Mod. Phys. \textbf{73}, 307 (2001).

\bibitem {details}Technical details will be given elsewhere.

\bibitem {Blonder82}G.~E. Blonder \textit{et al.}, Phys. Rev. B \textbf{25},
4515 (1982).

\bibitem {Fattori08}M.~Fattori \textit{et al.}, Phys. Rev. Lett. \textbf{100},
080405 (2008); M.~Gustavsson \textit{et al.}, ibid. \textbf{100}, 080404 (2008).

\bibitem {Metcalf99}H. J. Metcalf and P. van der Straten, \emph{Laser Cooling
and Trapping }(Springer-Verlag, New York, 1999).
\end{thebibliography}
\end{document}